\documentclass[floatfix,aps,prb,twocolumn,showpacs,superscriptaddress]{revtex4}
\usepackage{graphicx}
\usepackage[]{epsfig}
\usepackage{times,amsmath,amssymb}
\begin{document}
\title{Detection of spin polarization utilizing singlet and triplet states in a single-lead quantum dot}

\author{Tomohiro Otsuka}
\email[]{t-otsuka@meso.t.u-tokyo.ac.jp}
\affiliation{Department of Applied Physics, University of Tokyo, 7-3-1 Hongo, Bunkyo-ku, Tokyo 113-8656, Japan}

\author{Yuuki Sugihara}%
\affiliation{Department of Applied Physics, University of Tokyo, 7-3-1 Hongo, Bunkyo-ku, Tokyo 113-8656, Japan}

\author{Jun Yoneda}%
\affiliation{Department of Applied Physics, University of Tokyo, 7-3-1 Hongo, Bunkyo-ku, Tokyo 113-8656, Japan}

\author{Shingo Katsumoto}%
\affiliation{Institute for Solid State Physics, University of Tokyo, 5-1-5 Kashiwanoha, Kashiwa, Chiba 277-8581, Japan}
\affiliation{Institute for Nano Quantum Information Electronics, University of Tokyo, 4-6-1 Komaba, Meguro-ku, Tokyo 153-8505, Japan}

\author{Seigo Tarucha}%
\affiliation{Department of Applied Physics, University of Tokyo, 7-3-1 Hongo, Bunkyo-ku, Tokyo 113-8656, Japan}
\affiliation{Quantum-Phase Electronics Center, University of Tokyo, 7-3-1 Hongo, Bunkyo-ku, Tokyo 113-8656, Japan}
\affiliation{Institute for Nano Quantum Information Electronics, University of Tokyo, 4-6-1 Komaba, Meguro-ku, Tokyo 153-8505, Japan}

\date{\today}
\begin{abstract}
We propose and demonstrate a new method to probe local spin polarization in semiconductor micro devices at low and zero magnetic fields.
By connecting a single-lead quantum dot to a semiconductor micro device and monitoring electron tunneling into singlet and triplet states in the dot, we can detect the local spin polarization formed in the target device.
We confirm the validity of this detection scheme utilizing spin-split quantum Hall edge states.
We also observe nonzero local spin polarization at the device edge in low magnetic fields, which is not detectable with conventional macroscopic probes.
\end{abstract}

\maketitle


Spintronics, which utilizes the freedom of not only electron charge but also spin, has attracted strong interest in recent years.~\cite{2001WolfSci, 2004ZuticRMP}
In spintronics, generation and manipulation of spin polarization in nonmagnetic semiconductor microdevices are important challenges.
There are many theoretical proposals for generation~\cite{2001KiselevAPL, 2004PareekPRL, 2005EtoJPSJ, 2005OhePRB, 2008AharonyPRB} and manipulation~\cite{1990DattaAPL} of spin polarization utilizing spin-orbit interaction in semiconductor microdevices.~\cite{1955DresselhausPR, 1960RashbaFTT, 1997NittaPRL, 2002KogaPRL}
In recent years, related experiments for generation~\cite{2009DebrayNatPhys} and manipulation~\cite{2009FrolovNat} have been reported.
To understand the details of the mechanisms and operations, it is important to measure the local spin polarization formed in the real micro devices directly.
We need a local spin probe, which can access the local spin state of conducting electrons in semiconductor devices.

We have demonstrated a kind of local spin probe utilizing a single-lead quantum dot (SLQD).~\cite{2007OtsukaJPSJ, 2009OtsukaPRB}
We could detect the local spin polarization formed in a semiconductor quantum wire in high magnetic fields with minimal disturbance by monitoring tunneling of electrons into the spin-dependent Zeeman-split levels.
The usage of this method was limited, however, only to magnetic fields higher than 10 T because of the need for large Zeeman splitting.
We could not apply it directly to interesting spin phenomena such as generation of spin polarization utilizing spin-orbit interaction and spin polarization in the 0.7 anomaly of a quantum point contact (QPC),~\cite{1996ThomasPRL, 2000KristensenPRB} which occurs in lower or zero magnetic fields.
To overcome this problem, we here propose and demonstrate a scheme utilizing singlet and triplet states in an SLQD, which can be used in lower magnetic fields.


\begin{figure}[b]
\begin{center}
  \includegraphics{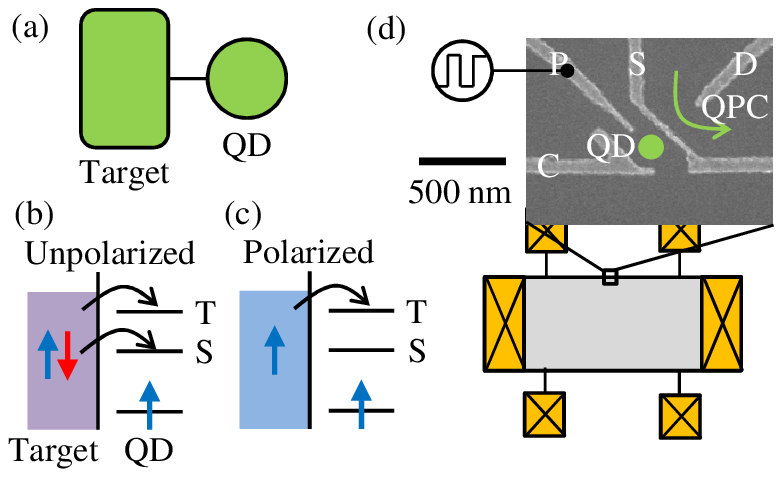}
  \caption{(Color online) (a) Schematic of the detector.
  A quantum dot is coupled to the target through a single tunneling barrier.
  (b),(c) Energy diagrams of the unpolarized case (b) and polarized case (c).
  In the unpolarized case, electrons tunnel into singlet and triplet states in the quantum dot.
  In the polarized case, tunneling into singlet state is suppressed.
  (d) Schematic of the device structure.
  At the edge of a Hall bar, a single-lead quantum dot is fabricated.
}
  \label{Scheme}
\end{center}
\end{figure}

To probe spin polarization of conduction electrons in a semiconductor microdevice, we couple an SLQD to the target device [Fig.~\ref{Scheme}(a)]  and measure tunneling of electrons into two-electron states in the SLQD.
The two-electron state is either a spin singlet or triplet state.
The singlet state is the ground state and the triplet state is the excited state in low magnetic fields.
Tunneling into these states reflects electron spin polarization in the target.
If the electrons in the target are spin unpolarized, we can observe tunneling into both singlet and triplet states in the SLQD [Fig.~\ref{Scheme}(b)].
On the other hand if the electrons are polarized, the tunneling into the singlet state is suppressed because this process needs an electron with opposite spin, which is not present in the target [Fig.~\ref{Scheme}(c)].
Thus we can get the information of the spin polarization from the measurement of tunneling rates into the singlet and triplet states.

This method has two important properties: good locality and small disturbance.
In the tunneling events, only the electrons which are close to the tunnel barrier can contribute to the process.
This gives good locality to the probe on the order of a few tens of nm.
By adopting the SLQD structure, we can reduce the leakage path of electrons through the probe to outer environments compared with the conventional quantum dot with two leads.
This results in extremely small disturbance in the measurement.
These properties are essential in the measurement of the fragile local spin polarization formed in semiconductor micro devices.
Also the SLQD structure is simpler than the conventional structure with two leads.
This simplicity contributes to the realization of smaller probes with higher yields.

In this Rapid Communication, first we confirm the operation of this detection scheme using spin-split edge states in the quantum Hall regimes.
The edge states are known to show well-defined spin polarization in relatively small magnetic fields and therefore can be used to check the operation of our detection scheme.
Due to the locality of our probe, we will detect the spin polarization of the outermost edge channel.
Second, with our new detection method, we investigate the detail of the local spin polarization of the outer edge states before the formation of well-defined spin-split channels, of which sign has been observed in the measurement of spin-dependent transport through a lateral quantum dot with the conventional two leads.~\cite{2001CiorgaPhysE}


Figure~\ref{Scheme}(d) shows a schematic of our device.
The device was fabricated from a GaAs/AlGaAs heterostructure wafer with sheet carrier density of 3.0~$\times$~10$^{15}$~m$^{-2}$ and mobility of 38~m$^2$/Vs.
We patterned a Hall bar mesa by wet-etching and deposited Ti/Au Schottky gates.
The SLQD was formed and coupled to the Hall bar by applying negative voltages on gates C, S, and P.
A QPC charge sensor to monitor the number of electrons in the SLQD, $N$, was formed by using gate D.

We measured the tunneling of electrons into the SLQD by monitoring the synchronized current through the nearby QPC charge sensor $I_{\rm sync}$ with square wave voltage excitation applied on gate P.~\cite{2004ElzermanAPL, 2008OtsukaAPL}
Upon this excitation, the chemical potential of the SLQD is periodically shifted up and down.
If the chemical potential of the target is in the range of this shift, electrons shuttle between the target and the SLQD by tunneling.
Because the electron incident into the SLQD raises an electrostatic potential in the surrounding, this synchronous shuttling decreases $I_{\rm sync}$, the value of which without shuttling is determined by direct electrostatic coupling between gate P and the QPC.
We can thus detect the electron tunneling as the decrease of  $I_{\rm sync}$.


\begin{figure}
\begin{center}
  \includegraphics{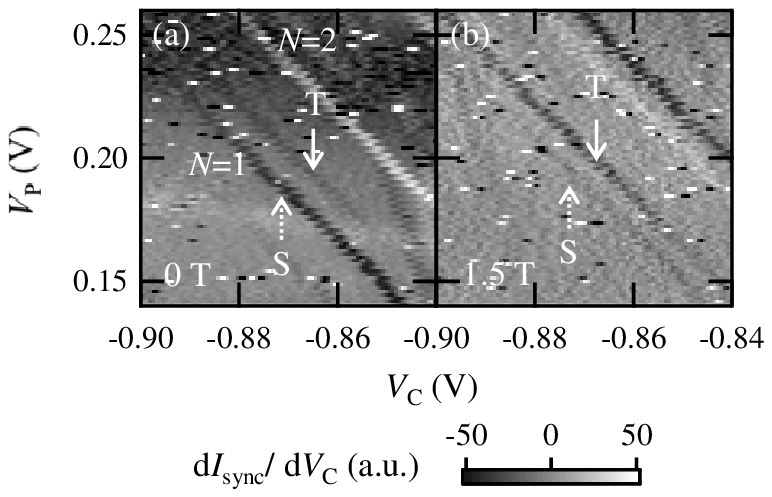}
  \caption{${\rm d}I_{\rm sync}/{\rm d}V_{\rm C}$ as a function of $V_{\rm P}$ and $V_{\rm C}$ at 0 T (a) and 1.5 T (b).
  Lower left (upper right) region corresponds to $N=1$ ($N=2$) and the dip structure in $I_{\rm sync}$ reflects the electron tunneling into two-electron states. 
  Up (down) arrows indicate the transition lines of the singlet (triplet) state.
}
  \label{Suppress}
\end{center}
\end{figure}

Figure~\ref{Suppress}(a) is the measured result of $I_{\rm sync}$ as a function of the voltages on gate P, $V_{\rm P}$, and C, $V_{\rm C}$, at zero magnetic field.
The amplitude of the square wave was set as 50~mV$_{\rm pp}$ to make the shift of the energy smaller than the charging energy but larger than singlet-triplet energy separation.
For clarity, we plot the numerical derivative ${\rm d}I_{\rm sync}/{\rm d}V_{\rm C}$.
The lower left area corresponds to the $N=1$ region and the upper right is the $N=2$ region.
A dip structure bounded by black and white lines between the $N=1$ and 2 region reflects the electron tunneling into the two-electron states of the SLQD. 
The dark lines in the dip structure correspond to the positions at which electrons start to tunnel into a specific state of the SLQD.
We can observe two dark lines.
The left one corresponds to tunneling into the ground singlet state [Fig~\ref{Suppress}(a) up arrow] and the right one reflects tunneling into the excited triplet state [Fig~\ref{Suppress}(a) down arrow].
The intensity of the singlet signal is significantly stronger.
This result indicates that the tunneling into the singlet is not forbidden when the spin polarization is not formed in the target at zero magnetic field.

Next, we applied the magnetic field of 1.5 T to form spin-split edge channels and checked the change of the electron tunneling.
The result of ${\rm d}I_{\rm sync}/{\rm d}V_{\rm C}$ is shown in Fig.~\ref{Suppress}(b).
The dark line corresponding to tunneling into the singlet state is faint compared to the triplet line, indicating that the tunneling into the singlet is suppressed.
This implies that the spin polarization is sufficiently formed in the outermost edge channel in the target and meets our detection scheme.
Note that the singlet-triplet transition was not observed up to 2 T in this device and our detection scheme was effective in this magnetic field range.
The duration of the square wave was set around 450 $\mu $s and this value is shorter than the expected spin relaxation time ($\sim $ms)~\cite{2005HansonPRL} in similar kinds of quantum dot systems, so we did not observe any effect which originates from spin relaxation.

In order for this scheme to work, the initial spin, which stays in the SLQD before the injection of the second spin,  should reflect the spin polarization of the target.
This requirement is satisfied in our experiment because the initial spin is refreshed by the target before its spin relaxation.
In the ejection phase, in which the level in the SLQD is shifted up by the square wave excitation, one of the two electrons goes out from the SLQD.
In this process, the initial electron can be ejected with a finite probability, for example 50\% in the case of the singlet.~\cite{comment1}
Then the initial electron will most likely be ejected in a few cycles before the spin relaxation, and the remaining electron, which becomes the initial electron in the next injection phase, reflects the spin polarization of the target.


From the measurement of $I_{\rm sync}$, we can extract the information of the tunneling rate into the SLQD $\Gamma $.
$\Gamma $ is related to the change of $I_{\rm sync}$ as
\begin{equation}
\label{gamma}
\Delta I_{\rm sync}= A \left( 1-\frac{\pi ^2}{\Gamma ^2\tau ^2 +\pi ^2} \right),
\end{equation}
where $A$ and $\tau $ are constants reflecting the sensitivity of the charge sensor and the duration of the square wave excitation respectively.~\cite{2004ElzermanAPL}
By using this relation, we can extract the tunneling rate into the singlet, $\Gamma _{\rm S}$, and triplet, $\Gamma _{\rm T}$.
$\Gamma _{\rm S}$ and $\Gamma _{\rm T}$ are represented by using the spin polarization of the target, $P$, as
\begin{equation}
\label{polarization1}
\Gamma _{\rm S}=B_{\rm S}(1+P)(1-P)
\end{equation}
\begin{equation}
\label{polarization2}
\Gamma _{\rm T}=B_{\rm T}(P^2+3),
\end{equation}
where $B_{\rm S}$ and $B_{\rm T}$ are constants reflecting the effective tunneling barriers of the singlet and triplet states respectively.
Here we assumed that the spin state of the first electron in the SLQD was proportional to $P$.
By using these equations, we can extract $P$ from the measurement of $I_{\rm sync}$.~\cite{comment2}

\begin{figure}
\begin{center}
  \includegraphics{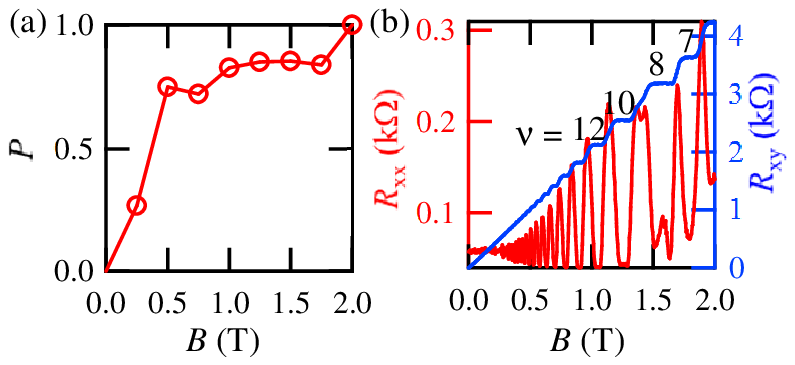}
  \caption{(Color online) (a) $P$ as a function of $B$ measured by the local probe utilizing a single-lead quantum dot.
  Nonzero $P$ is observed even at the magnetic field as low as 0.5 T.
(b) Measurement of $R_{\rm xx}$ and $R_{\rm xy}$ as a function of $B$ using conventional macroscopic voltage probes in the same device to (a).
The labels show values of the filling factor $\nu $.
Spin splitting is observed only in magnetic fields higher than 1.3 T.
}
  \label{Polarization}
\end{center}
\end{figure}

Figure~\ref{Polarization}(a) shows the extracted $P$ as a function of the magnetic field $B$.
In this analysis, we minimized the effect of the change of $\Gamma $ with $B$ by adjusting $V_{\rm C}$ to keep $\Gamma _{\rm S}+\Gamma _{\rm T} $ constant. Also we assumed that the value of $P$ at zero magnetic field was 0.~\cite{comment4}
The obtained $P$ increases with $B$ and finally reaches $P\approx 1$ at around 2~T.
This reflects the formation of spin-split edge channels at the edge of the Hall bar.
By considering the low $R_{\rm xx}$ at $\nu =7$, almost perfect spin polarization will be formed in magnetic fields larger than 1.7~T and our local probe shows $P\approx 1$ in this magnetic field range.
This shows the validity of our detection scheme.

Another important feature in this figure is the relatively high spin polarization observed at a magnetic field as low as 0.5 T.
In such a low magnetic field, we usually do not expect formation of spin-split edge channels in standard transport measurement, but this is not the case for our local probe measurement.
This result is also consistent with the previous experiment of the spin-dependent transport through a lateral quantum dot with the two-dimensional electron gas leads.~\cite{2001CiorgaPhysE}


Figure~\ref{Polarization}(b) shows the measurement of longitudinal resistance $R_{\rm xx}$ and transverse resistance $R_{\rm xy}$ as a function of $B$ with conventional macroscopic voltage probes in the same device.
$R_{\rm xx}$ shows conventional Shubnikov-de Haas oscillations. 
In the magnetic field range larger than 1.3 T, there appear additional minimum in $R_{\rm xx}$ and plateaus in $R_{\rm xy}$.~\cite{comment3}
This means that the spin-split edge channels are only visible for $B>1.3$ T with the macroscopic probes, whereas the finite spin polarization is already visible for $B>0.5$ T with our microscopic probes.
This difference makes clear the advantage of our SLQD detection to locally probe the spin polarization.

A possible reason for this difference can be the local spin polarization.
The measured objects of the microscopic and macroscopic probes are different.
The microscopic probe using the SLQD detects the local spin polarization formed at the device edge.
On the other hand, the macroscopic probes using conventional voltage probes detect the separation of spin-split Landau levels, i.e. the spatial separation of spin-split edge channels.
In very low magnetic fields, there is no spin polarization and edge states in the target and spin polarization is not observed with both probes.
In the intermediate magnetic fields, separation of spin split edge channels will not be perfect but there will be spatial distribution of spin polarization in the Hall bar.
With the macroscopic probes, we cannot detect the spin polarization.
With the microscopic local probes, we can detect the local spin polarization at the device edge.
By considering the result in Fig.~\ref{Polarization}(a), the local spin polarization at the device edge will increase rapidly around 0.5 T and saturate to the maximum value.
In high magnetic fields larger than 1.3~T, spatially separated spin-split edge channels come to be formed and the macroscopic probes start to show finite spin polarization.


In conclusion, we have proposed a new scheme utilizing two-electron states in a single-lead quantum dot to probe local spin polarization formed in semiconductor microdevices in low and zero magnetic fields.
We have confirmed the work of our detection scheme utilizing quantum Hall edge states as the target.
We have extracted the information of spin polarization and detected finite spin polarization at the magnetic field as low as 0.5 T, which could not be detected with conventional macroscopic voltage probes.
This detection technique will be applicable to explore interesting spin phenomena in low magnetic fields such as generation of spin polarization with spin-orbit interaction and spin polarization in the 0.7 anomaly of a quantum point contact.

We thank T. Nakajima, T. Obata, and G. Allison for fruitful discussions and technical support.
This work was supported by a Grant-in-Aid for Scientific Research and Special Coordination Funds for Promoting Science and Technology.


\begin{references}
\bibitem{2001WolfSci}
S. A. Wolf, D. D. Awschalom, R. A. Buhrman, J. M. Daughton, S. von Moln\'ar, M. L. Roukes, A. Y. Chtchelkanova, and D. M. Treger,
Science {\bf 294}, 1488 (2001).

\bibitem{2004ZuticRMP}
I. \v{Z}uti\'c, J. Fabian, and S. Das Sarma,
Rev. Mod. Phys. {\bf 76}, 323 (2004).

\bibitem{2001KiselevAPL}
A. A. Kiselev, and K. W. Kim,
Appl. Phys. Lett. {\bf 78}, 775 (2001).

\bibitem{2004PareekPRL}
T. P. Pareek,
Phys. Rev. Lett. {\bf 92}, 076601 (2004).

\bibitem{2005EtoJPSJ}
M. Eto, T. Hayashi, and Y. Kurotani,
J. Phys. Soc. Jpn. {\bf 74}, 1934 (2005).

\bibitem{2005OhePRB}
J. Ohe, M. Yamamoto, T. Ohtsuki, and J. Nitta,
Phys. Rev. B {\bf 72}, 041308(R) (2005).

\bibitem{2008AharonyPRB}	
A. Aharony, O. Entin-Wohlman, Y. Tokura and S. Katsumoto, 
Phys. Rev. B {\bf 78}, 125328 (2008).

\bibitem{1990DattaAPL}	
S. Datta and B. Das, 
Appl. Phys. Lett. {\bf 56}, 665 (1990).

\bibitem{1960RashbaFTT}
E. I. Rashba,
Fiz. Tverd. Tela (Leningrad) 2, 1224 (1960)
[Sov. Phys. Solid State {\bf 2}, 1109 (1960)].

\bibitem{1955DresselhausPR}
G. Dresselhaus,
Phys. Rev. {\bf 100}, 580 (1955).

\bibitem{1997NittaPRL}
J. Nitta, T. Akazaki, H. Takayanagi and T. Enoki,
Phys. Rev. Lett. {\bf 78}, 1335 (1997).

\bibitem{2002KogaPRL}
T. Koga, J. Nitta, T. Akazaki and H. Takayanagi,
Phys. Rev. Lett. {\bf 89}, 046801 (2002).

\bibitem{2009DebrayNatPhys}	
P. Debray, S. M. S. Rahman, J. Wan, R. S. Newrock, M. Cahay, A. T. Ngo, S. E. Ulloa, S. T. Herbert, M. Muhammad and M. Johnson, Nature Nanotechnol. {\bf 4}, 759 (2009).

\bibitem{2009FrolovNat}
S. M. Frolov, S. Luscher, W. Yu, Y. Ren, J. A. Folk and W. Wegscheider,
Nature (London) {\bf 458}, 868 (2009).

\bibitem{2007OtsukaJPSJ}
T. Otsuka, E. Abe, S. Katsumoto, Y. Iye, G. L. Khym, and K. Kang,
J. Phys. Soc. Jpn. {\bf 76}, 084706 (2007).

\bibitem{2009OtsukaPRB}
T. Otsuka, E. Abe, Y. Iye and S. Katsumoto, 
Phys. Rev. B {\bf 79}, 195313 (2009).

\bibitem{1996ThomasPRL}
K. J. Thomas, J. T. Nicholls, M. Y. Simmons, M. Pepper, D. R. Mace and D. A. Ritchie,
Phys. Rev. Lett. {\bf 77}, 135 (1996).

\bibitem{2000KristensenPRB}
A. Kristensen, H. Bruus, A. E. Hansen, J. B. Jensen, P. E. Lindelof, C. J. Marckmann, J. Nyg\aa rd, C. B. S\o rensen, F. Beuscher, A. Forchel and M. Michel, Phys. Rev. B {\bf 62}, 10950 (2000).

\bibitem{2001CiorgaPhysE}
M. Ciorga, A. S. Sachrajda, P. Hawrylak, C. Gould, P. Zawadzki, Y. Feng and Z. Wasilewski,
 Physica E {\bf 11}, 35 (2001).

\bibitem{2004ElzermanAPL}
J. M. Elzerman, R. Hanson, L. H. W. van Beveren, L. M. K. Vandersypen, and L. P. Kouwenhoven,
Appl. Phys. Lett. {\bf 84}, 4617 (2004).

\bibitem{2008OtsukaAPL}
T. Otsuka, E. Abe, Y. Iye, and S. Katsumoto,
Appl. Phys. Lett. {\bf 93}, 112111 (2008).

\bibitem{2005HansonPRL}
R. Hanson, L. H. W. van Beveren, I. T. Vink, J. M. Elzerman, W. J. M. Naber, F. H. L. Koppens, L. P. Kouwenhoven and L. M. K. Vandersypen,
Phys. Rev. Lett. {\bf 94}, 196802 (2005).

\bibitem{comment1}
Here we expected that the effect of spin selectivity of the edge states in the ejection phase is small.
This would be caused by the charge redistribution in the edge states,\cite{1992ChklovskiiPRB} which makes the interedge distance above the Fermi level shorter than the distance below the level.
In the measurement with square wave excitation, the state in the SLQD is shifted far above the Fermi level on the order of meV in the ejection phase.
Then the interedge distance is small and the spin selectivity will be negligible.

\bibitem{comment2}
For qualitative discussion, we assumed the tunneling rates in the injection phase $\Gamma _{\rm in}$ and the ejection phase $\Gamma _{\rm out}$ are the same in this paper.
But more precisely, $\Gamma _{\rm in}$ and $\Gamma _{\rm out}$ might be different. 
These values will be evaluated by using real-time measurement of the tunneling events.~\cite{2004ElzermanNat}

\bibitem{comment4}
First, we evaluated $\Gamma _{\rm S}/\Gamma _{\rm T}$ using Eq. (1).
The unknown parameter $A$ was obtained  from $\Delta I _{\rm sync}$ with small negative $V_{\rm C}$, because $\Gamma \tau \gg \pi $ and Eq.~(1) becomes $\Delta I_{\rm sync}=A$.
Then we evaluated $P$ using $\Gamma _{\rm S}/\Gamma _{\rm T}$, Eqs. (2) and (3).
The unknown parameter $B_{\rm S}/B_{\rm T}$ was evaluated from the data at the zero magnetic field $(\Gamma_{\rm S}/\Gamma_{\rm T})_{0\rm T}=B_{\rm S}/3B_{\rm T}$ and assumed to have no magnetic field dependence.

\bibitem{comment3}
A peak structure in $R _{\rm xx}$ around 1.55 T is a fake peak caused by an imperfect Ohmic contact.

\bibitem{1992ChklovskiiPRB}
D. B. Chklovskii, B. I. Shklovskii, and L. I. Glazman,
Phys. Rev. B {\bf 46}, 4026 (1992).

\bibitem{2004ElzermanNat}
J. M. Elzerman, R. Hanson, L. H. Willems van Beveren, B. Witkamp, L. M. K. Vandersypen and L. P. Kouwenhoven, 
Nature (London) {\bf 430}, 431 (2004).

\end{references}
\end{document}